\documentclass[aps,pra,twocolumn,groupedaddress,showpacs,showkeys,superscriptaddress]{revtex4-1}

\usepackage{amsmath}
\usepackage{amssymb}
\usepackage{graphicx}
\usepackage{epstopdf}

\usepackage{subcaption}
\usepackage{mwe}

\usepackage{media9}
\usepackage{hyperref}

\begin{document}

\title{Propagation of Strongly Hybridized Edge Modes in Plasmonic Waveguides}
\thanks{This short paper is based on the material presented at META-2019 conference. It is a prequel to our work \cite{pshenichnyuk-2019} published later.}

\author{I.A. Pshenichnyuk}
\email[correspondence address: ]{i.pshenichnyuk@skoltech.ru}
\affiliation{Skolkovo Institute of Science and Technology, Moscow 121205, Russian Federation}

\author{S.S. Kosolobov}
\affiliation{Skolkovo Institute of Science and Technology, Moscow 121205, Russian Federation}

\author{A.I. Maimistov}
\affiliation{National Research Nuclear University MEPhI (Moscow Engineering Physics Institute), Moscow 115409, Russian Federation }

\author{V.P. Drachev}
\affiliation{Skolkovo Institute of Science and Technology, Moscow 121205, Russian Federation}
\affiliation{University of North Texas, Denton, Texas 76203, USA}

\begin{abstract}
In our work we investigate the propagation of optical modes in nanoscale hybrid plasmonic waveguides. Frequency domain Maxwell equations based simulations are implemented to study properties of mixed modes in 3D. The results of our analysis are applied to design an advanced hybrid waveguide with useful characteristics. It is demonstrated here how to couple two edge plasmons with a waveguide mode to form a hybrid waveguide well tuned for applications in the field of electro-optical transistors (modulators). A mixed polarization state of edge plasmons allows them to couple with a horizontally polarized (TE) optical mode (which is not possible for usual surface plasmons) and open the way to polarization independent plasmonic modulators.
\end{abstract}

\maketitle

\section{Introduction} \label{intro}

Surface plasmon polariton (SPP) modes provide a subwavelength optical confinement of light and play a significant role in nano-optics. Other important characteristics of these modes include the enhanced field intensity and strong light-matter interaction (via the oscillations of surface electrons plasma). SPPs are extremely useful in photonics and allow to construct small and efficient devices. For example, they can be used in combination with epsilon-near-zero (ENZ) switching to design compact waveguide based electro-optical modulators \cite{sorger-2012}. Hybridization of light and matter in the form of SPP quasiparticles takes place even at the level of single photons \cite{kolesov-2009}. It allows to talk about potential applications of SPPs in future quantum plasmonic circuits \cite{tame-2013,deleon-2012} and molecular electronics \cite{pshenichnyuk-2011,pshenichnyuk-2013}. In general, the usage of hybrid quasiparticles and their combinations appears as a powerful paradigm providing the ground for future photonics. Another vivid example is related to exciton-polaritons that demonstrate many interesting and useful properties. Their ability to form Bose condensates at room temperature is discussed in literature \cite{plumhof-2014}. Being condensed, exciton-polaritons demonstrate nonlinear coherent behavior typical for quantum fluids \cite{pshenichnyuk-2017,pshenichnyuk-2018,pshenichnyuk-2015}. Various applications of exciton-polaritons in photonics are already suggested \cite{berloff-2017,liew-2010}.

One well-known drawback of SPP modes is related to losses. Finite propagation length of plasmons limits the usage of extremely compact SPP waveguides in large scale photonic circuits. On the other hand, integrated dielectric waveguides provide almost loss-free way to transmit light, but the confinement is poor (to compare with the typical sizes of modern electronic components). The idea of hybrid plasmonic waveguides (HPWG) is to mix two approaches and obtain a compromise solution with a desired proportion between losses and confinement \cite{alam-2014}.

In classical hybrid waveguides \cite{dai-2009} a layer of metal, separated by a thin dielectric, is placed on top of an ordinary waveguide. In this case at least two hybrid modes are formed and participate in the evolution of the mixture. Even in the simples case, strong coupling between a waveguide mode and plasmonic mode does not allow to apply the coupled waves theory to describe the evolution and require numerical calculations.

One principle limitation of hybrid waveguides is related to the polarization of plasmons. SPPs interact only with waveguide modes polarized perpendicular to the metallic surface. HPWG with more sophisticated structure are needed for applications that require an advanced polarization handling. Other important parameters, like the waveguide mode to plasmon conversion length and conversion losses define the practical applicability of HPWG.

In this work we propose a model of HPWG where a dielectric waveguide mode is hybridized with two edge SPP modes. Edge plasmons with a mixed polarization state are used in our design to couple otherwise orthogonal modes to each other. During the propagation of the strongly hybridized mixture we observe a smooth low loss conversion of the waveguide mode into a plasmon and back accompanied by a polarization transformation. It is discussed how to use the proposed conversion principle to expand the idea of a waveguide based plasmonic modulator and remove polarization restrictions typical for this class of devices. Numerically heavy 3D calculations are performed to model the proposed HPWG.

\section{Discussion} \label{discuss}

\begin{figure}
\centerline{\includegraphics[width=0.47\textwidth]{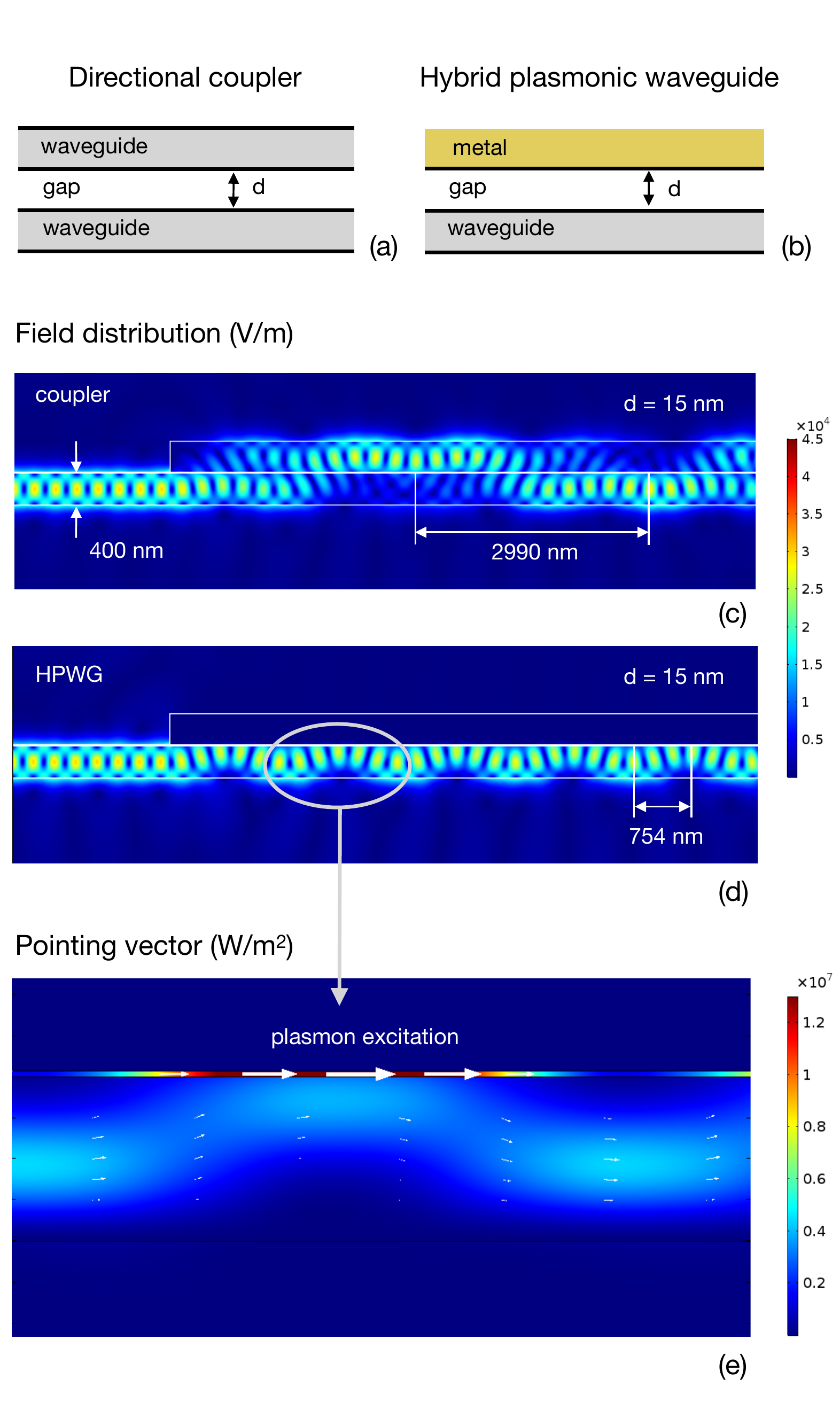}}
\caption {Propagation of a hybrid mode in a directional coupler and a simple plasmonic waveguide. (a) Scheme of a directional coupler with two waveguides separated by the gap $d$. (b) Simplest possible model of a hybrid plasmonic waveguide containing one waveguide and one plasmon guiding metallic surface. (c) Propagation of light in a coupler is accompanied by a periodic transfer of population between two waveguides (d) Similar process takes place in a hybrid waveguide but the period of transformation, coupling between modes and, in general, nature of the interacting modes is different. (e) Pointing vector is plotted to demonstrate in details the act of transformation of a waveguide mode into plasmon and back. \label{fig_hpwg}}
\end{figure}

Propagation of coupled waves is usually accompanied by periodic oscillations of mode populations. There is a certain analogy between the behavior of HPWG and directional couplers, formed by two dielectric waveguides placed close to each other (see Fig.~\ref{fig_hpwg}a). The propagation of the electric field inside the coupler is plotted in Fig.~\ref{fig_hpwg}c. The optical signal travels back and forth between two waveguides. To obtain simplest possible HPWG we exchange one of the waveguides by a metallic surface supporting SPP modes (Fig.~\ref{fig_hpwg}b). The propagation of a hybrid mode in this case is accompanied by the partial conversion of the waveguide mode into plasmon and back (Fig.~\ref{fig_hpwg}b) that resembles mode conversion in directional couplers. Unfortunately, the theory of coupled waves, well developed for couplers is not efficient for HPWG, because of a strong interaction \cite{alam-2014}. To study the propagation of strongly coupled plasmonic modes numerical calculations are necessary.

Plasmonic modes are extremely compact and allow to reach large values of the field intensity, which is useful for many applications. In the example presented here (Fig.~\ref{fig_hpwg}), plasmons are formed in $15$ nm gap between the dielectric waveguide and the metallic surface. The details of one conversion are shown in Fig.~\ref{fig_hpwg}e where the Pointing vector is plotted (color corresponds to the absolute value and the direction is shown by arrows). Since SPP modes are lossy, it is problematic to build purely plasmonic large scale circuits. Hybrid waveguides allow to reach a compromise between confinement and losses. The waveguide mode may be transformed to plasmon only when it is required (for example, when the interaction with an electronic part of the circuit is necessary) and then transformed back again.

\begin{figure}
\centerline{\includegraphics[width=0.47\textwidth]{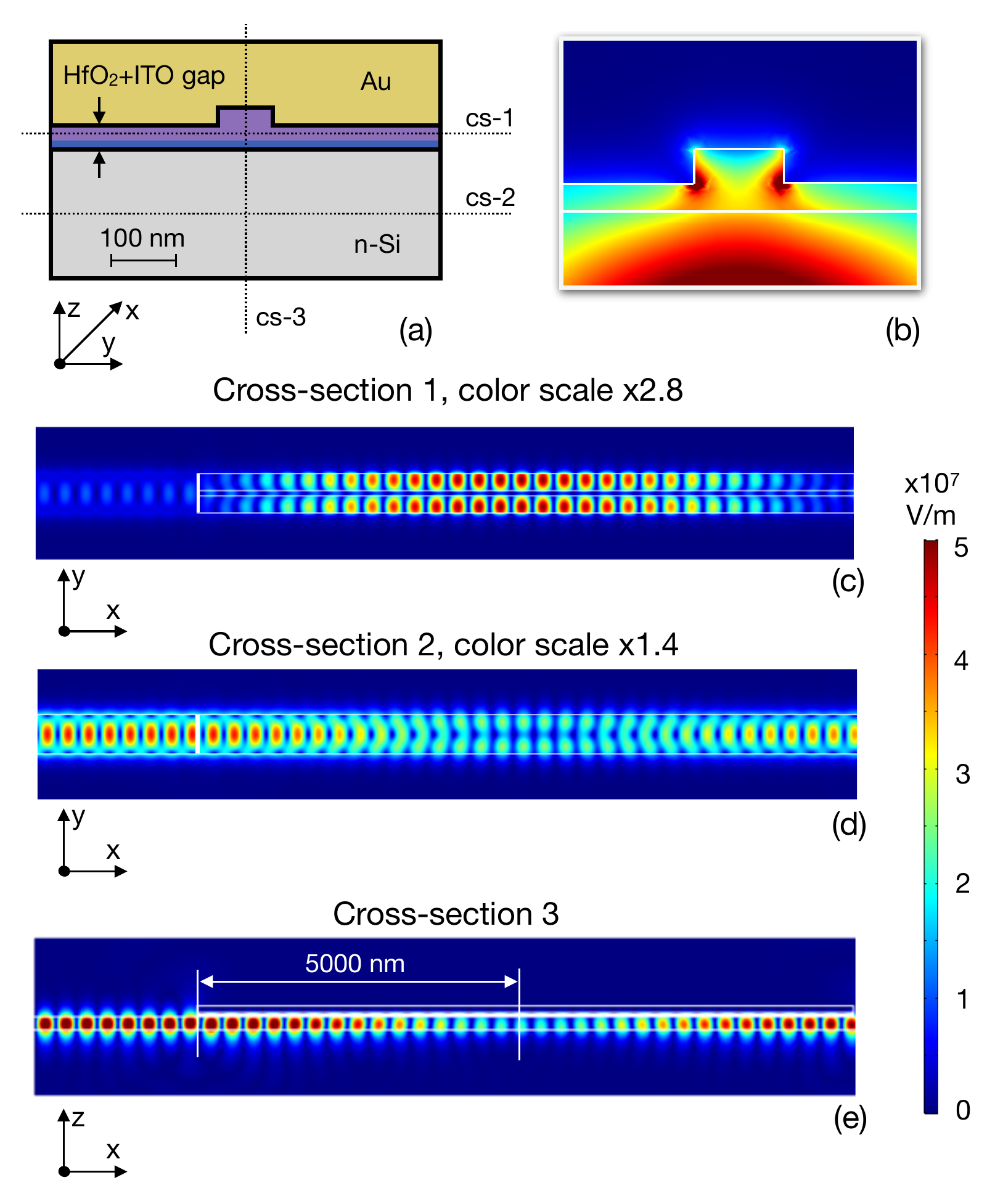}}
\caption {The model of a hybrid waveguide mixing two edge plasmons with an ordinary waveguide mode. (a) Transverse cross-section of the waveguide, demonstrating the structure of the sandwich and implemented materials. (b) Electric field distribution in the transverse cross-section. The waveguide mode is coupled with two edge plasmons. (c),(d),(e) Absolute value of the electric field distribution inside the structure. Different cross-sections are demonstrated. \label{fig_edgehybrid}}
\end{figure}

The development of advanced HPWG for applications in photonics requires the control of various parameters including the plasmon conversion length and conversion efficiency. In general, each transformation causes certain intensity losses. For this reason one should avoid multiple transformations (as, for example, in Fig.~\ref{fig_hpwg}d) and the length of the device should be adjusted accordingly. The transformation should be also smooth, to avoid significant insertion losses. To provide low loss input and output, HPWG modes effective indices (at least one of them) should be close to the effective index of the waveguide mode. At the same time, the length of the device should be large enough to perform a required operation. For example, the comfortable length for plasmon assisted electro-optical modulation is of the order of $10$~$\mu$m \cite{sorger-2012}. The transformation length usually depends on coupling between modes and in the case of plasmonic waveguides it is large. Practically it is hard to reach the desired level of coupling just varying the size of the gap between the waveguide and metallic surface, since the high precision control of the distance is required. The precise control of coupling is, thus, another challenge during the development of HPWG.

Another major issue is related to polarization restrictions. SPP interact only with TM waveguide modes. This fact limits the universality of HPWG based devices.
Here we propose a 3D model of advanced HPWG designed for telecom wavelength $1550$ nm, that mixes a waveguide mode with two edge plasmons (see Fig.~\ref{fig_edgehybrid}b). Our model addresses all practical questions formulated above. The transverse cross-section of the device is shown in Fig.~\ref{fig_edgehybrid}a. On top of a single mode silicon waveguide (designed for TE polarization) we place a 3-layer sandwich. It consists of  a $10$ nm layer of dielectric (HfO$_2$, dark blue), $15$ nm layer of indium tin oxide (ITO, violet) and $150$ nm thick gold layer. The refractive indices of ITO and the dielectric are close to each other ($n\,{\approx}\,2$) and the boundary between them is almost optically homogeneous. The HfO$_2$/ITO junction is provided for applications, where the switching behavior is required (like, for example, modulators). Applying the voltage one may initiate ENZ effect in ITO which leads to the strong attenuation of the signal \cite{sorger-2012}. 
Above the ITO layer we place an additional block of ITO, which we call a plasmonic rail. The metallic electrode placed on top of the structure forms two gold edges at the Au/ITO boundary that support edge plasmonic modes (see Fig.~\ref{fig_edgehybrid}b).
Edge plasmons have a mixed polarization state and may couple with the TE mode of the waveguide.
The width and height of the plasmonic rail are two additional geometrical parameters which can be used to control the coupling strength.
Since the rail support two plasmons we obtain a 3 component hybrid mode (Fig.~\ref{fig_edgehybrid}b). The geometrical defect (plasmonic rail) brakes a symmetry in $y$-direction and makes 2D modelling of HPWG insufficient.

The propagation of a hybrid mode is shown in Fig.~\ref{fig_edgehybrid}c-e where the absolute value of the electric field is plotted. Three different cross-sections of the 3D model are shown, two horizontal and one vertical, which can be considered as 'top view', 'bottom view' and 'side view'. The exact position of cross-sections are shown by dotted lines in Fig.~\ref{fig_edgehybrid}a. In the 'bottom view' the horizontal plane is fixed at the middle of the waveguide which allows to see mainly the waveguide mode. In the 'top view' the horizontal plane passes through the center of the gap between the waveguide and the metal and allows to see the plasmons.

The figure demonstrates how the TE-polarized waveguide mode passing through the sandwich smoothly  transforms into a couple of edge plasmons and then returns back into the waveguide (Fig.~\ref{fig_edgehybrid}d-f). Smooth conversion is achieved by the matching of effective indices of the waveguide mode and HPWG. Losses on double conversion including the insertion losses are close to $1.5$~dB.
The waveguide to plasmon transformation length in our particular model is $5$~$\mu$m. The sandwich double that size is long enough to perform efficient electro-optical modulation using the principle described in \cite{sorger-2012}. Such a sandwich contains just one conversion period, which helps to keep plasmonic losses under control. If HPWG is not terminated at the right length, the second transformation starts immediately after the first one.

The conversion length depends on coupling and can be tuned using the geometrical parameters (width and height) of the plasmonic rail. It is important to stress that the rail is the only coupling provider in our model. In the absence of this element the horizontally polarized TE waveguide mode does not couple to the plasmon. Our coupling mechanism also allows to make plasmonic coupling weaker and the conversion length longer. Instead of one parameter responsible for the coupling in classical HPWG (gap size), we get two geometrical parameters (width and height of the rail).
During the conversion the polarization state of light partially changes from TE to TM and back. Ordinary surface plasmons can not be used for such a conversion. The implemented principle can be also used to construct compact polarization convertors \cite{pshenichnyuk-2018c}.

\begin{figure}
\centerline{\includegraphics[width=0.47\textwidth]{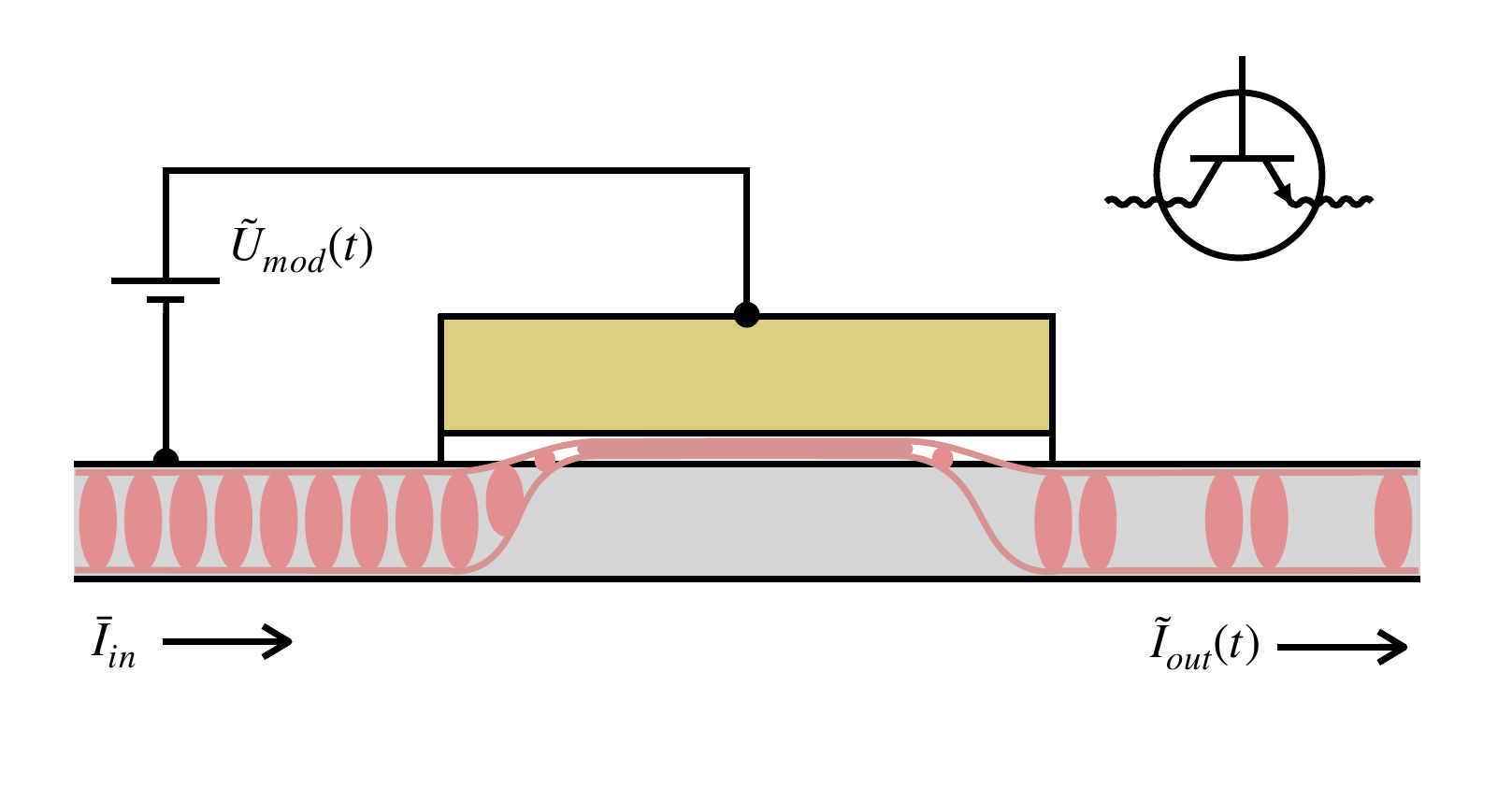}}
\caption {Schematic illustration of a hybrid plasmonic waveguide based electro-optic transistor (modulator). Passing through the modulating sandwich, light shrinks into plasmon. In this state it can be efficiently modulated and transformed back to the waveguide mode. \label{fig_modulator}}
\end{figure}

Convenient (and adjustable) transformation length, relatively low losses and gentle treatment of polarization  makes the designed nanostructure useful for the construction of efficient electro-optical transistors or modulators without usual plasmonic polarization restrictions. The concept is illustrated in Fig.~\ref{fig_modulator}. Being converted to a plasmon, optical signal becomes highly compressed and intensified. In such a state it is well suited for the efficient modulation. As it is described above, our design supports a thin layer of insulator between the silicon waveguide and ITO, that forms a capacitor. Applying an external voltage to the top metallic layer (it also plays a role of an electrode) one may cause the charge accumulation at the boundary of the dielectric. It is well known that certain density of charges accumulated in ITO may initiate ENZ effect. In this state the refractive index changes significantly providing an efficient switching of the modulator \cite{sorger-2012}.

\section*{Conclusion}

In our work we analyze the propagation of a hybrid mode formed by the combination of two plasmonic edge modes and a waveguide mode. The plasmonic rail is incorporated into our geometry to maintain the coupling between modes (otherwise uncoupled). Numerically expensive 3D computations are employed in the analysis. The parameters of the structure are chosen to provide smooth transformation of the waveguide mode into the plasmon and back with minimal losses. The conversion length can be tuned modifying the geometry of the rail. The suggested vertically assembled structure allows to couple TE-polarized waveguide modes with SPP modes. It is proposed to use the designed HPWG to produce efficient waveguide based plasmonic modulators without polarization restrictions.


%

\end{document}